\documentclass[aps, prd, amsmath, floats, floatfix, twocolumn, nofootinbib, superscriptaddress, showpacs]{revtex4-1}
\usepackage{graphicx,color}
\usepackage{latexsym}
\usepackage{amsmath,amsfonts,amssymb}
\usepackage{physics}     

\def\bN{{\mathbb N}}

\begin{document}


\title{Superradiant instabilities in the Kerr-mirror and Kerr-AdS black holes \\ with Robin boundary conditions}


\author{Hugo R. C. Ferreira}
\affiliation{Istituto Nazionale di Fisica Nucleare -- Sezione di Pavia, Via Bassi 6, 27100 Pavia, Italy}

\author{Carlos A. R. Herdeiro}
\affiliation{Departamento de F\'isica da Universidade de Aveiro and Center for Research and Development in Mathematics and Applications (CIDMA), Campus de Santiago, 3810-183 Aveiro, Portugal}


\date{December 2017}

\begin{abstract}
It has been recently observed that a scalar field with Robin boundary conditions (RBCs) can trigger both a superradiant and a bulk instability for a BTZ black hole (BH)~\cite{Dappiaggi:2017pbe}. To understand the generality and scrutinize the origin of this behavior, we consider here the superradiant instability of a Kerr BH confined either in a mirror-like cavity or in AdS space, triggered also by a scalar field with RBCs. These boundary conditions are the most general ones that ensure the cavity/AdS space is an isolated system, and include, as a particular case, the commonly considered Dirichlet boundary conditions (DBCs). Whereas the superradiant modes for some RBCs differ only mildly from the ones with DBCs, in both cases we find that as we vary the RBCs, the imaginary part of the frequency may attain arbitrarily large positive values. We interpret this growth as being sourced by a bulk instability of both confined geometries when certain RBCs are imposed to either the mirror-like cavity or the AdS boundary, rather than by energy extraction from the BH, in analogy with the BTZ behavior.
\vspace*{7ex}
\end{abstract}

\maketitle

\section{Introduction}

The phenomenon of superradiance in black hole (BH) physics is a fascinating and intriguing \textit{classical} process through which energy can be extracted from the BH. Over the last few decades, a considerable body of work has been devoted to understanding the many faces of this phenomenon (see~\cite{Brito:2015oca} for a review). In particular, a thorough understanding of superradiant \textit{instabilities} has proved to be challenging. Such instabilities arise when superradiant modes are trapped around the BH and may undergo repeated amplifications~\cite{Press:1972zz}. The instability that ensues can be seen at the linear level, but probing its full development requires a non-linear dynamical analysis (see~\cite{Sanchis-Gual:2015lje,Bosch:2016vcp,Sanchis-Gual:2016tcm,East:2017ovw,Herdeiro:2017phl} for fully non-linear treatments and their interpretation in various setups). 

There are three commonly used trapping mechanisms to induce superradiant instabilities of a \textit{scalar} field around a Kerr BH: (1) the existence of a mass term, first observed in~\cite{Damour:1976kh} and further developed in~\cite{Detweiler:1980uk,Zouros:1979iw,Strafuss:2004qc,Konoplya:2006br,Dolan:2007mj,Dafermos:2008en,Rosa:2009ei,Cardoso:2011xi,Hod:2012zza,Dolan:2012yt,Shlapentokh-Rothman:2013ysa,East:2013mfa,Brito:2014wla,Hod:2016iri}; (2) imposing the field is confined in a cavity around the BH, via a trapping boundary condition, which was the initial BH bomb proposal~\cite{Press:1972zz}, and further developed in~\cite{Cardoso:2004nk,Hod:2014pza,Hod:2016rqd}; (3) imposing AdS asymptotics~\cite{Cardoso:2004hs}. The first type of mechanism determines the asymptotic scalar field solution to be an exponentially decaying mode, with frequency smaller (in modulus) than the field mass. But for mechanisms (2) and (3) there is  freedom in the choice of the field's behavior at the mirror/AdS boundary. Commonly, one chooses Dirichlet Boundary Conditions (DBCs), which guarantee that the system is isolated and there is no energy-momentum flux through the boundary. However, the same is guaranteed by Robin Boundary Conditions (RBCs), a more generic choice which has not yet been thoroughly studied in the literature. Exceptions include some recent studies within the (2+1)-dimensional BTZ BH \cite{Ferreira:2017cta,Dappiaggi:2017pbe} and some studies for spin 1 fields~\cite{Wang:2015goa,Wang:2015fgp}. 

The study of a massive scalar field with RBCs in a rotating BTZ BH~\cite{Dappiaggi:2017pbe} showed that, unlike for DBCs, a superradiant instability can occur, whose signature are some exponentially growing modes that extract energy from the BH. However,  it was also observed in~\cite{Dappiaggi:2017pbe}  that not \textit{all} of the modes which grow exponentially in time extract energy from the BH. This behavior was interpreted as due to an intrinsic, bulk instability of the AdS space, which occurs for certain RBCs, as previously noted in \cite{Dappiaggi:2016fwc}. There is then an interplay between the superradiance instability, which takes place for moderately low frequencies, and the bulk instability, which happens for the lowest frequencies.

The occurrence of this bulk instability is associated to the use of (a subset of) RBCs. A natural question is how generic it is and, in particular, if it is strictly associated to the AdS geometry, or if it occurs for other spacetimes mimicking the global AdS geometry, such as a volume of Minkowski spacetime in which the scalar field is confined due to a mirror-like boundary condition imposed at the boundary of this volume. Contextualizing, one may recall that the well known turbulent instability of AdS~\cite{Bizon:2011gg} also occurs for a scalar field confined in the Minkowski-mirror setup~\cite{Maliborski:2012gx}.

In this paper, we show that an analogous bulk instability occurs in the case of a massive scalar field in a Kerr-AdS BH with RBCs at infinity. In order to be able to obtain analytical results, we consider some approximations, similarly to \cite{Cardoso:2004hs}: a small, slowly rotating Kerr-AdS BH and scalar modes with wavelengths much larger than the typical size of the BH. Our results show that, as we vary the RBCs at infinity such that the real part of the frequency tends to zero, the imaginary part of the frequency becomes arbitrarily large and, as a consequence, this bulk instability becomes dominant. 

Performing a similar analysis for the Kerr-mirror system, on the other hand, in which we impose RBCs at the mirror's location, we find that, besides mild modifications to the superradiance effect studied in \cite{Cardoso:2004nk} for the Kerr-mirror system with DBCs, there is also a new unstable mode for certain choices of RBCs, similarly to the Kerr-AdS case. This suggests that the new mode instability is caused when certain RBCs are imposed at either the mirror in the Kerr-mirror system or the AdS boundary in the Kerr-AdS system, and that it is not fundamentally associated to the asymptotic structure of these confining geometries.


The content of the present paper is as follows. In Section~\ref{sec:RBCs} we review the RBCs to be employed at either the mirror or the AdS infinity of the systems under consideration. In Section~\ref{sec:Kerr-mirror}, we obtain the superradiant modes for the Kerr-mirror system with RBCs imposed at the mirror's location, whereas in Section~\ref{sec:Kerr-AdS} we repeat the computation for the Kerr-AdS BH with RBCs imposed at infinity. We present the conclusions in Section~\ref{sec:conclusions}. Throughout the paper we employ natural units in which $c = G_{\rm N} = 1$ and a metric with signature $(-+++)$.

\section{Robin boundary conditions}
\label{sec:RBCs}

In addressing the phenomenon of superradiance, one must guarantee that the amplification of the field is sourced by the BH, and not some other energy source. Thus one requires that the field-BH system is isolated. As explicitly shown in the particular case of the (2+1)-dimensional BTZ BH in \cite{Ferreira:2017cta,Dappiaggi:2017pbe}, this is achieved when one considers generic RBCs, for either a BH-mirror or BH-AdS system.

Consider for concreteness the case of a scalar field in a Kerr-AdS BH, as described in detail in Section~\ref{sec:Kerr-AdS}. Therein, we construct two linearly independent mode solutions of the Klein-Gordon equation, $\Phi^{\rm (D)}(t,r,\theta,\phi)$ and $\Phi^{\rm (N)}(t,r,\theta,\phi)$. $\Phi^{\rm (D)}$ is chosen to be the \textit{principal solution} at $r \to \infty$, that is, the unique solution (up to scalar multiples) such that $\lim_{r \to \infty} \Phi^{\rm (D)}(t,r,\theta,\phi)/\Psi(t,r,\theta,\phi) = 0$ for every solution $\Psi$ that is not a scalar multiple of $\Phi^{\rm (D)}$. This is the Dirichlet solution. The other solution, $\Phi^{\rm (N)}$, is a nonprincipal solution and it is not unique. We shall call it the Neumann solution. 

A general solution may be written as a linear combination of these two solutions. In order for the energy flux to be zero at infinity, we require the scalar field to satisfy RBCs, in which case the solution is written as
\begin{equation} \label{eq:solutionRBC}
\Phi = \mathcal{N} \left[ \cos(\zeta) \Phi^{\rm (D)} + \sin(\zeta) \Phi^{\rm (N)} \right] \, ,
\end{equation}
where $\zeta \in [0,\pi)$ parametrizes the RBCs and $\mathcal{N}$ is a normalisation constant. This is the form we shall use in the following sections. Observe that the DBCs correspond to $\zeta=0$, whereas the Neumann boundary conditions (NBCs) correspond to $\zeta=\frac{\pi}{2}$.

Furthermore, a second feature of the RBCs is that the solution \eqref{eq:solutionRBC} must be square-integrable near infinity,
\begin{equation}
\int^{\infty} \dd r \sqrt{-g} \, g^{tt} \left|R(r)\right|^2 < \infty \, .
\end{equation}
Using the results of Section~\ref{sec:Kerr-AdS}, we can show that in the AdS case this restricts the range of the mass parameter $\mu^2$ of the scalar field such that $-\frac{9}{4}<\mu^2<-\frac{5}{4}$. We note, however, that in the case of the BH-mirror system, in which the RBCs are imposed at the mirror at a finite radial distance, there is no such restriction in the mass of the scalar field and, thus, for simplicity, we will consider the massless case in Section~\ref{sec:Kerr-mirror}.

\section{The Kerr-mirror system}
\label{sec:Kerr-mirror}

In this section we shall consider a massless scalar field in the Kerr-mirror system with the RBCs at the mirror's location. We follow closely the computation in~\cite{Cardoso:2004nk}, but to keep this paper self-contained we provide some details of the computation. 

The metric of a Kerr BH is given in Boyer-Lindquist coordinates by
\begin{multline}
\dd s^2 = - \frac{\Delta-a^2 \sin^2 \theta}{\rho^2} \dd t^2 - \frac{4 M r a \sin^2 \theta}{\rho^2} \dd t \dd \phi + \frac{\rho^2}{\Delta} \dd r^2 \\
+ \rho^2 \dd \theta^2 + \sin^2 \theta \frac{(r^2+a^2)^2 - \Delta a^2 \sin^2 \theta}{\rho^2} \dd \phi^2 \, ,
\end{multline}
where
\begin{equation*}
\Delta \equiv r^2 + a^2 - 2Mr \, , \qquad \rho^2 \equiv r^2 + a^2 \cos^2 \theta \, ,
\end{equation*}
$M$ is the BH mass and $J\equiv Ma$ its angular momentum.

We consider a massless scalar field $\Phi$ that satisfies the Klein-Gordon equation $\nabla^2 \Phi = 0$ in the background of the Kerr BH. Taking the ansatz,
\begin{equation}
\Phi(t,r,\theta,\phi) = e^{-i \omega t + i m \phi} S_l^m(\theta) R(r) \, ,
\end{equation}
where $S_l^m(\theta)$ are the spheroidal angular functions, $\omega \geqslant 0$ is the field's frequency and $m\in \mathbb{Z}$ is the azimuthal harmonic index. The Klein-Gordon equation separates~\cite{Brill:1972xj} into the angular equation for $S_l^m(\theta)$,
\begin{equation}
\frac{\partial_{\theta} \left(\sin \theta \, \partial_{\theta} S_l^m \right)}{\sin \theta} + \left[ a^2 \omega^2 \cos^2 \theta - \frac{m^2}{\sin^2 \theta} + A_{lm} \right] S_l^m = 0 \, ,
\end{equation}
and the radial equation for $R$,
\begin{multline}
\Delta \partial_r \left(\Delta \partial_r R \right) + \big[ \omega^2 (r^2 + a^2)^2 - 2 M a m \omega r \\
+ a^2 m^2 - \Delta (a^2 \omega^2 + A_{lm}) \big] R = 0 \, ,
\end{multline}
where $A_{lm}$ is the separation constant, which reduces to the standard $l(l+1)$ in the Schwarzschild limit. Expansions for this constant in terms of the spheroidicity parameter can be found in~\cite{Seidel:1988ue}.

\subsection{Superradiant modes by a matching method}


Assuming a small frequency and slow rotation approximation, $M \ll 1/\omega$ and $a \ll M$, the exterior region to  the event horizon is divided into two regions, wherein the radial equation is solved separately. One then matches the near region solution, valid for $r-r_+ \ll 1/\omega$, with the far region solution, valid for $r-r_+ \gg M$, in an overlapping part of both regions $M \ll r-r_+ \ll 1/\omega$, which always exists for sufficiently small frequencies.

\subsubsection{Near region $r-r_+ \ll 1/\omega$}


In the near region the radial equation is given by
\begin{equation} \label{eq:nearradialeq}
	\Delta \frac{\dd}{\dd r} \left(\Delta \frac{\dd R}{\dd r}\right) + r_+^4 (\omega - m \Omega_{\mathcal{H}})^2 R - l(l+1) R = 0 \, .
\end{equation}
Introducing a new radial coordinate
\begin{equation}
	z \equiv \frac{r-r_+}{r-r_-} \, , \quad 0 \leqslant z < 1 \, ,
\end{equation}
and defining
\begin{equation}
	R(z) \equiv z^{i \varpi} (1-z)^{l+1} F(z) \, ,
\end{equation}
with $\varpi \equiv \frac{r_+^2}{r_+-r_-}(\omega-m\Omega_{\mathcal{H}})$, the radial equation may be written as
\begin{equation}
	z(1-z)\frac{\dd^2 F}{\dd z^2} ´\left[c-(a+b+1)z\right] \frac{\dd F}{\dd z} -ab F = 0 \, ,
\end{equation}
with
\begin{equation}
	a \equiv l+1-i2\varpi \, , \quad
	b \equiv l+1 \, , \quad
	c \equiv 1 + i 2 \varpi \, .
\end{equation}
This is the Gaussian hypergeometric equation \cite{NIST} and the most general solution for $R$ may be written as
\begin{align}
	R(z) &= (1-z)^{l+1} \left[A z^{-i \varpi} F(a-c+1,b-c+1;a-c;z)  \right. \notag \\
	&\quad + \left. B z^{i \varpi} F(a,b;c;z) \right] \, .
	\label{req12}
\end{align}
$A,B$ are two integration constants. The first term in~\eqref{req12} represents an ingoing mode at the horizon, while the second term represents an outgoing mode at the horizon. For the study of quasinormal modes/quasibound states (they are equivalent in this context), we set $B=0$.

The large $r$ behavior of the near region solution may be obtained by making the transformation $z \to 1-z$ \cite{NIST},
\begin{align*}
	R(z) &= z^{-i \varpi} \Bigg[\frac{\Gamma(2-c)\Gamma(a+b-c)}{\Gamma(a-c+1)\Gamma(b-c+1)} \, (1-z)^{-l} \\
	&\quad \times F(1-a,1-b;c-a-b+1;1-z) \\
	&\quad +  \frac{\Gamma(2-c)\Gamma(c-a-b)}{\Gamma(1-a)\Gamma(1-b)} \, (1-z)^{l+1}  \\
	&\quad \times F(a-c+1,b-c+1;a+b+1-c;1-z) \Big] \, ,
\end{align*}
As $z \to 1$ and $1-z \to (r_+-r_-)/r$,
\begin{align} \label{e:largernear}
	R(r) &\approx A \Gamma(1-i2\varpi) \left[\frac{(r_+-r_-)^{-l} \Gamma(2l+1)}{\Gamma(l+1) \Gamma(l+1-i2\varpi)} r^l \right. \notag \\
	&\quad \left. + \frac{(r_+-r_-)^{l+1} \Gamma(-2l-1)}{\Gamma(-l) \Gamma(-l-i2\varpi)} r^{-l-1} \right] \, .
\end{align}
This is Eq.~(20) in~\cite{Cardoso:2004nk}.

\subsubsection{Far region $r-r_+ \gg M$}


In the far region, the radial equation may be approximated by the wave equation of a scalar field mode of frequency $\omega$ and angular momentum $l$ in flat spacetime,
\begin{equation}
	\frac{\dd^2 (rR)}{\dd r^2} + \left[\omega^2 - \frac{l(l+1)}{r^2}\right](rR) = 0 \, .
\end{equation}
The most general solution is
\begin{equation} \label{eq:solutionsphereicalBessel}
	R(r) = \alpha j_{l}(\omega r) + \beta y_{l}(\omega r) \, ,
\end{equation}
where $j_l$ and $y_l$ are the spherical Bessel functions of the first and second kind \cite{NIST}, respectively, and $\alpha,\beta$ are two constants. 
We place a mirror in the exterior region at $r=r_0$ wherein RBCs are imposed, of the form, $cf.$~\eqref{eq:solutionRBC},
\begin{equation} \label{eq:RBCmirror}
	\cos(\zeta) R(r_0) + \sin(\zeta) R'(r_0) = 0 \, , \quad \zeta \in [0, \pi) \, .
\end{equation}
where the prime stands for derivative with respect to the argument.
This fixes the ratio
\begin{equation} \label{eq:betaalpha1}
	\frac{\beta}{\alpha} = - \frac{\cos(\zeta) j_l(\omega r_0) + \sin(\zeta) j'_l(\omega r_0)}{\cos(\zeta) y_l(\omega r_0) + \sin(\zeta) y'_l(\omega r_0)} \, ,
\end{equation}
which generalizes Eq.~(28) in~\cite{Cardoso:2004nk}.

The small $r$ behavior of the far region solution is
\begin{equation} \label{eq:smallrfar}
	R(r) \approx \alpha \frac{\omega^l}{(2l+1)!!} r^l - \beta (2l-1)!! \, \omega^{-l-1} r^{-l-1} \, . 
\end{equation}
\begin{figure*}[t]
	\centering
	\includegraphics[width=0.47\linewidth]{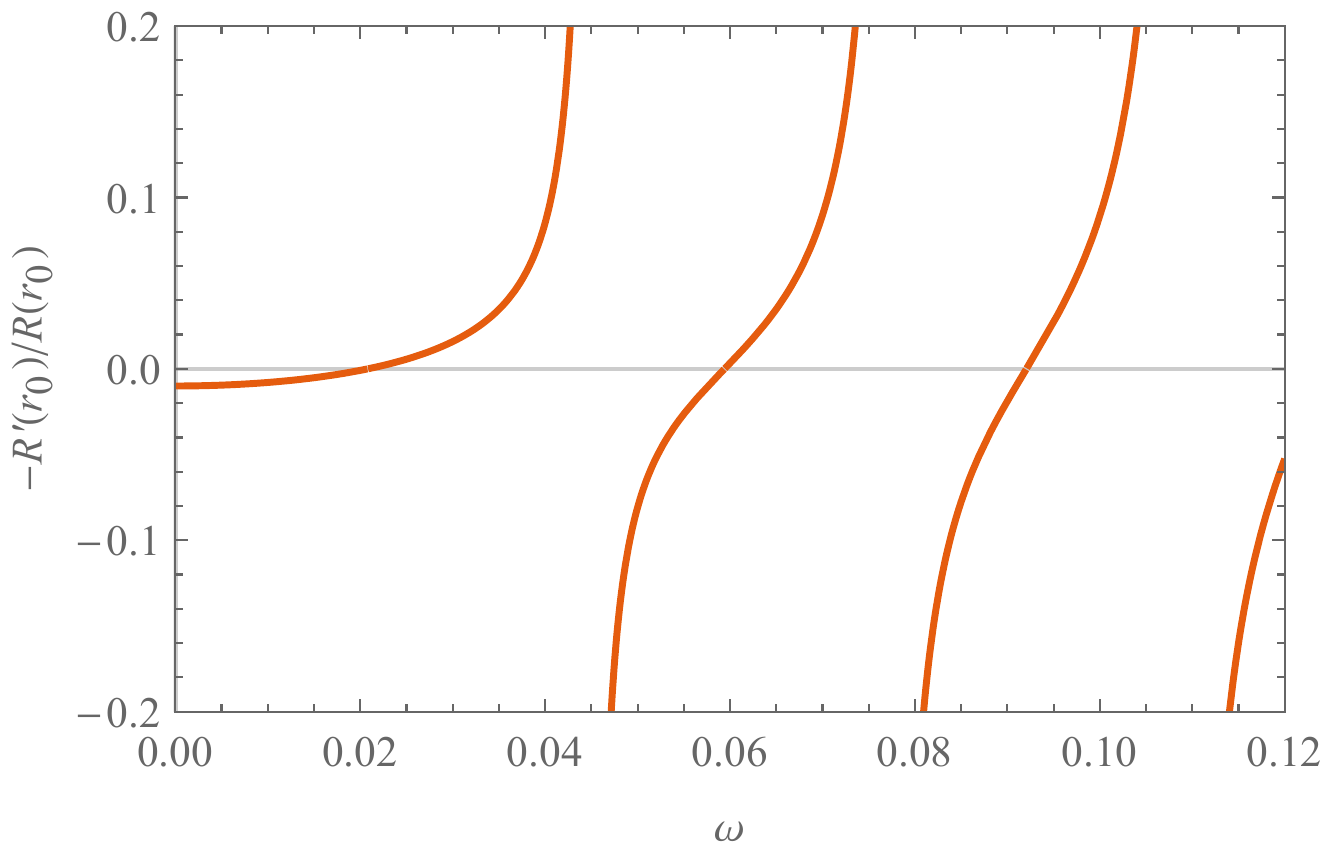} \hspace*{5ex}
	\includegraphics[width=0.47\linewidth]{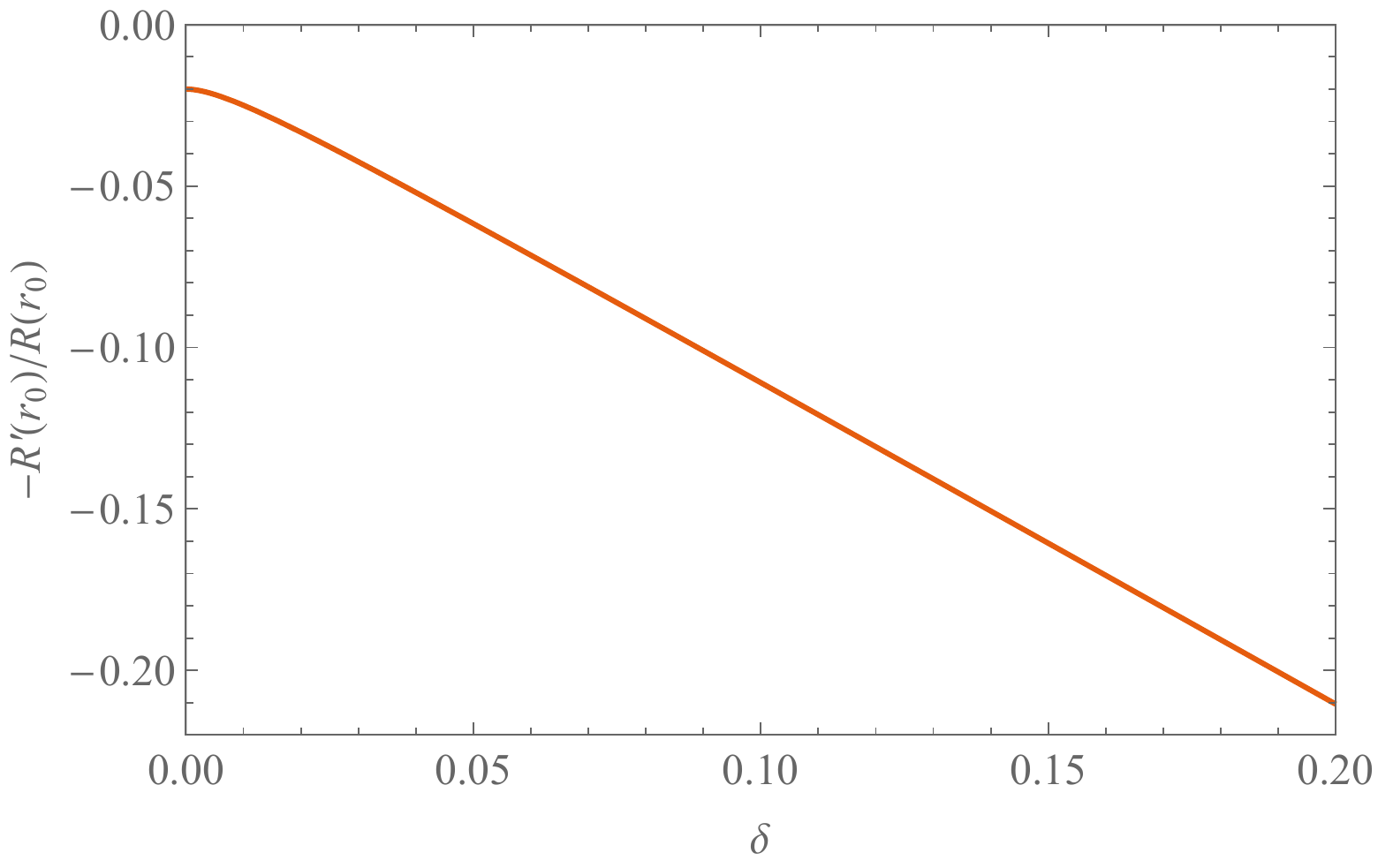}
	\caption{\label{fig:plot-Kerr-mirror-CD}Normal modes of a scalar field in a Minkowski-mirror cavity under RBCs. The plots show $-R'(r_0)/R(r_0) = \cot(\zeta)$ $vs.$ $\omega$ (left panel) and $vs.$ $\delta$ (right panel) for a cavity in Minkowski spacetime with $r_0=100$ and a mode with $l=1$. The real zeros of this function correspond to the Neumann normal frequencies, whereas the singularities correspond to the Dirichlet normal frequencies. The purely imaginary frequencies never occur for either Dirichlet or Neumann boundary conditions.
	\vspace*{3ex}}
\end{figure*}

\subsubsection{Matching in overlapping region $M \ll r-r_+ \ll 1/\omega$}

In the overlapping region $M \ll r-r_+ \ll 1/\omega$ one can match the large $r$ solution in the near region \eqref{e:largernear} with the small $r$ solution in the far region \eqref{eq:smallrfar}, giving
\begin{align}
	\frac{\beta}{\alpha} &= - \frac{(r_+-r_-)^{2l+1} \Gamma(l+1)\Gamma(-2l-1)}{\Gamma(2l+1)\Gamma(-l) (2l+1)!! (2l-1)!!} \notag \\
	&\quad \times \frac{\Gamma(l+1-i2\varpi)}{\Gamma(-l-i2\varpi)} \omega^{2l+1} \, .
\end{align}
This expression can be further simplified using the relation $\Gamma(z+1)=z\Gamma(z)$, giving
\begin{align} \label{eq:betaalpha2}
	\frac{\beta}{\alpha} &= - i 2 \varpi \, \frac{(r_+-r_-)^{2l+1}}{(2l)!(2l+1)(2l+1)!} \left(\frac{l!}{(2l-1)!!}\right)^2 \notag \\
	&\quad \times \left(\prod_{k=1}^{l} \left(k^2+4\varpi^2\right)\right) \omega^{2l+1} \notag \\
	&\equiv - i \Lambda \varpi \omega^{2l+1} \, .
\end{align}
Using this and \eqref{eq:betaalpha1}, we can find a relation between the RBC and the frequency
\begin{equation} \label{eq:Minkowskinormalmodes}
\cot(\zeta) = - \frac{R'(r_0)}{R(r_0)} = - \frac{j'_l(\omega r_0) + \frac{\beta}{\alpha} \ y'_l(\omega r_0)}{j_l(\omega r_0) + \frac{\beta}{\alpha} \ y_l(\omega r_0)} \, .
\end{equation}
In Fig.~\ref{fig:plot-Kerr-mirror-CD} we plot $\cot(\zeta)$ as a function of real frequency $\omega$ and purely imaginary $i\delta$ for the case $\beta/\alpha=0$, $i.e.$ no BH. These give the \textit{normal modes} of the scalar field in a cavity in Minkowski spacetime, under RBCs. 

Note that the purely imaginary frequency modes only exist for a \emph{subclass} of RBCs, which moreover does not include the DBCs and NBCs. This can be heuristically understood as follows. For purely imaginary frequencies, the solution \eqref{eq:solutionsphereicalBessel} has an exponential behavior in $r$, and as such both the solution and its derivative do not vanish at a given $r$. However, for certain values of $\zeta$, it is possible that a linear combination of the solution and its derivative vanishes, satisfying \eqref{eq:RBCmirror}. Analogous results were observed in \cite{Dappiaggi:2017pbe}.

\subsubsection{Approximation $\Im[\omega] \ll \Re[\omega] \ll 1$}

\begin{figure*}[t!]
	\centering
	\includegraphics[width=0.393\linewidth]{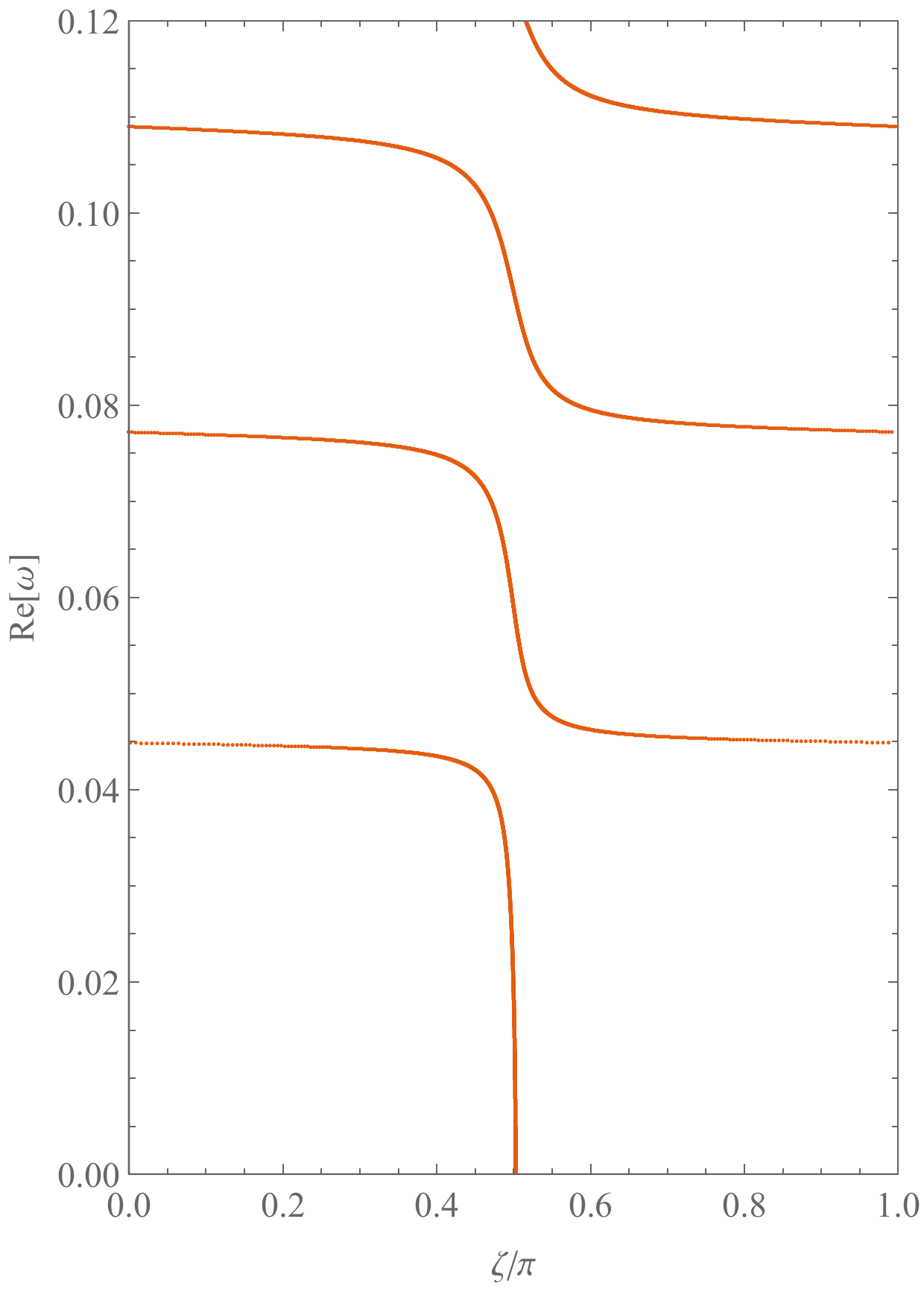} \hspace*{12ex}
	\includegraphics[width=0.385\linewidth]{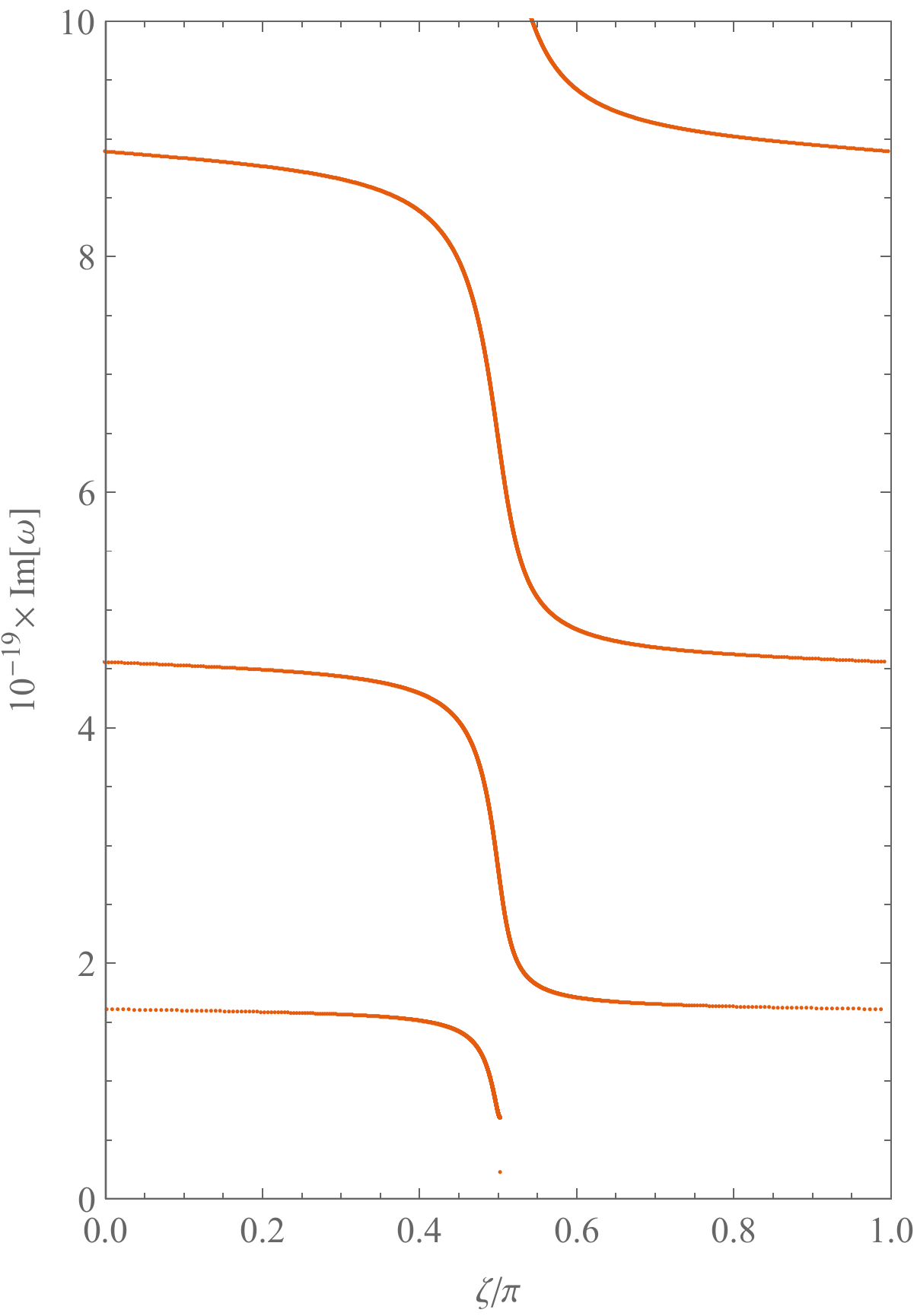}
	\caption{\label{fig:plot-Kerr-mirror-1} Superradiant modes in a Kerr-mirror cavity. Real part (coinciding with the real normal frequencies) and imaginary part of the frequencies as a function of $\zeta/\pi$ in the Kerr-mirror system with $M=0.01$, $a=0.001$, $r_0=100$, $l=1$ and $m=1$.
	\vspace*{3ex}
	}
\end{figure*}

We now consider an approximation appropriate to the problem at hand, $\Im[\omega] \ll \Re[\omega] \ll 1$. Under this approximation, we can compute the superradiant modes, which can also be considered quasinormal modes in this context, as a small deformation of the real normal modes. In this case, the RHS of \eqref{eq:betaalpha2} is very small and can be taken to be zero in first approximation. Then, \eqref{eq:betaalpha1} yields
\begin{equation}
	\cos(\zeta) j_l(\omega r_0) + \sin(\zeta) j'_l(\omega r_0) \approx 0 \, .
\end{equation}
Denote by $a_{l,n}^{\zeta}$, with $n \in \bN$, the real, positive roots of the above equation for $\omega r_0$. Then, in first approximation, the real part of the quasinormal frequencies may be given by $\Re[\omega_n] \approx a_{l,n}^{\zeta}/r_0$.

Now, write the full solution as $\omega_{\rm QN} \equiv a_{l,n}^{\zeta}/r_0 + i \tilde{\delta}/r_0$, where we assume that $\tilde{\delta} \ll a_{l,n}^{\zeta}$. One may write
\begin{equation*}
	\frac{\beta}{\alpha} = -i \tilde{\delta} \, \frac{\cos(\zeta) j'_l(a_{l,n}^{\zeta}) + \sin(\zeta) j''_l(a_{l,n}^{\zeta})}{\cos(\zeta) y_l(a_{l,n}^{\zeta}) + \sin(\zeta) y'_l(a_{l,n}^{\zeta})} + \order{\tilde{\delta}^2} \, .
\end{equation*}
Comparing with \eqref{eq:betaalpha2} gives
\begin{equation}
	\delta \equiv \frac{\tilde{\delta}}{r_0} \approx \varpi \frac{\cos(\zeta) y_l(a_{l,n}^{\zeta}) + \sin(\zeta) y'_l(a_{l,n}^{\zeta})}{\cos(\zeta) j'_l(a_{l,n}^{\zeta}) + \sin(\zeta) j''_l(a_{l,n}^{\zeta})} \frac{(a_{l,n}^{\zeta})^{2l+1}}{r_0^{2l+2}} \Lambda \, .
\end{equation}
Assuming that, for all $\zeta \in [0,\pi)$,
\begin{equation} \label{eq:negativeterm}
	\frac{\cos(\zeta) y_l(a_{l,n}^{\zeta}) + \sin(\zeta) y'_l(a_{l,n}^{\zeta})}{\cos(\zeta) j'_l(a_{l,n}^{\zeta}) + \sin(\zeta) j''_l(a_{l,n}^{\zeta})} < 0 \, ,
\end{equation}
then, $\delta > 0$ if $\varpi < 0$, that is, if $\Re[\omega_n] < m \Omega_{\mathcal{H}}$. This is the standard result for superradiance in asymptotically flat BH spacetimes, which generalizes the results obtain in \cite{Cardoso:2004nk} in the case of DBCs. Note that the sign of $\delta$ does not depend on the position of the mirror in the exterior region of the BH. In particular, whether the mirror is beyond the speed of light surface or not is irrelevant for the existence of superradiant modes.

It remains to show the inequality \eqref{eq:negativeterm} for all $\zeta \in [0,\pi)$. For $\zeta = 0$ it is known that $y_l(a_{l,n}^{0})/j'_l(a_{l,n}^{0}) < 0$ and we have numerical evidence that the inequality is satisfied for other values of $\zeta$.

Illustrative examples of the variation of the real and imaginary parts of the frequencies of superradiant modes are provided in Fig.~\ref{fig:plot-Kerr-mirror-1}. The figure shows that changing the value of $\zeta$ away from the DBCs value ($\zeta=0$) changes only  mildly the values obtained with the DBCs, decreasing it. Note that, given the approximation being employed, the accuracy of the numerical results decreases as $\omega$ approaches zero.

Also note that as $\omega \to 0$, one has that
\begin{equation}
\zeta \to \arccot\left[\frac{1}{r_0}\left(\frac{1}{2}-\frac{\Gamma \left(l+\frac{3}{2}\right)}{\Gamma \left(l+\frac{1}{2}\right)}\right)\right] \, .
\end{equation}
In particular, for very large $r_0$, as $\omega$ goes to zero $\zeta$ tends to $\frac{\pi}{2}$, corresponding to NBCs.

\subsubsection{Approximation $\Re[\omega] \ll \Im[\omega]$}

\begin{figure}[t!]
	\centering
	\includegraphics[width=0.84\linewidth]{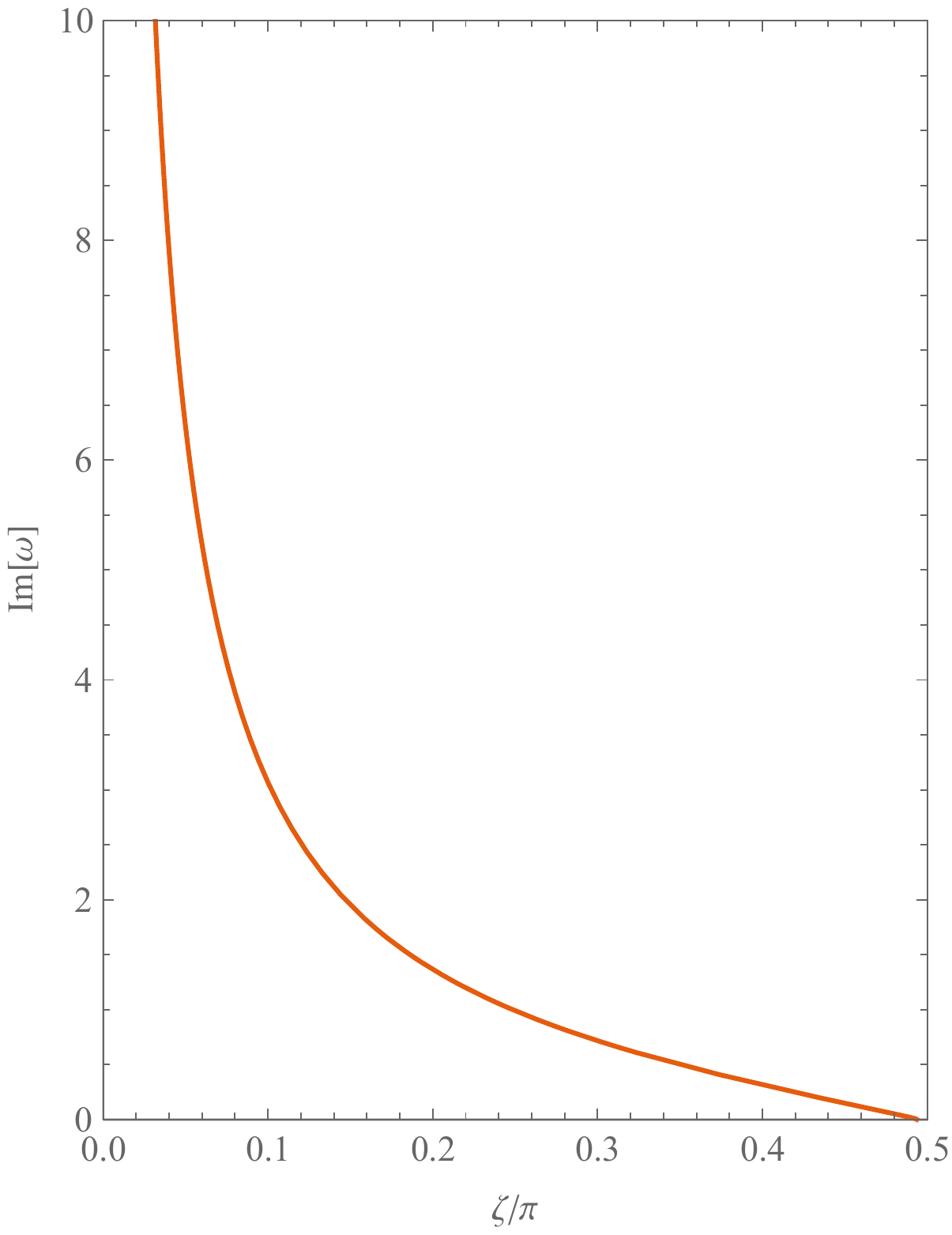}
	\caption{\label{fig:plot-Kerr-mirror-2} Purely imaginary \emph{normal} modes in the Minkowski-cavity system. The imaginary part of the frequency as a function of $\zeta/\pi$ in the Minkowski-cavity system with $r_0=100$ and $l=1$.}
\end{figure}

In the previous section, we computed the quasinormal frequencies in the Kerr-mirror cavity whose imaginary part is much smaller than the real part, treating these frequencies as ``perturbations'' of the real normal frequencies of the Minkowski-mirror cavity. However, the latter also has purely imaginary normal frequencies. Therefore, the BH system must also have quasinormal frequencies whose imaginary part is much greater than the real part and which can be treated as ``perturbations'' of purely imaginary frequencies of the Minkowski-mirror system.

The purely imaginary normal modes in the Minkowski-cavity system can be numerically computed by using \eqref{eq:Minkowskinormalmodes} with $\beta/\alpha = 0$. In Fig.~\ref{fig:plot-Kerr-mirror-2} we show the frequencies $i \delta$ with positive imaginary part for a representative example.

First, note that these modes decay exponentially as we move away from the mirror, but also grow exponentially in time. They are therefore \emph{unstable} modes, which only exist when a subset of Robin boundary conditions (which exclude both the Dirichlet and Neumann ones) are imposed at the mirror. Second, observe that, as $\zeta \to 0^+$, the imaginary part $\delta \to \infty$. That is, the growth rate of these modes can be arbitrarily large as the parameter $\zeta$ characterizing the RBCs gets arbitrarily small. This type of behavior has also been recently found in the context of AdS spacetimes (as developed in Section~\ref{sec:Kerr-AdS}) but, to the best of our knowledge, it has not been discussed for flat spacetimes with boundaries.

The next step in our calculation would be to obtain the approximate real part of the quasinormal frequencies in the Kerr-mirror system, which are much smaller than the corresponding imaginary part, in an analogous way to the previous case. However, in that case, the full approximation being employed was $\Im[\omega] \ll \Re[\omega] \ll 1$, which was necessary in order to use \eqref{eq:betaalpha2}. However, in the present case, we cannot assume that $\Re[\omega] \ll \Im[\omega] \ll 1$, since $\Im[\omega]$ will necessarily be arbitrarily large as $\zeta \to 0^+$. Therefore, the strategy to compute the quasinormal frequencies as ``perturbations'' of the purely imaginary frequencies of the Minkowski-mirror system, using the current approximation scheme, fails for these frequencies. As we will see in Section~\ref{sec:Kerr-AdS}, the same is not true for the Kerr-AdS system, for which we can obtain the approximate real part of the quasinormal frequencies.

Nevertheless, the above calculations are sufficient to draw some conclusions about the Kerr-mirror system. In the regime $\Im[\omega] \ll \Re[\omega] \ll 1$, the behavior for different RBCs is qualitatively similar to that for Dirichlet boundary conditions and there is superradiance when $\Re[\omega] < m \Omega_{\mathcal{H}}$. However, for a certain range of RBCs, which does not include the usual DBCs and NBCs, there is a mode with positive imaginary part of the frequency much greater than the real part. As explained further in the Kerr-AdS case, the flux of energy across the horizon for a given mode with frequency $\omega + i\delta$ is proportional to $\delta^2 + \omega(\omega - m \Omega_{\mathcal{H}})$ and, thus, for large enough $\delta$, the flux is \emph{towards} the BH, even if $\omega < m \Omega_{\mathcal{H}}$. This means that such a mode, even though is unstable, is \emph{not} extracting energy from the BH. This constitutes a \emph{new} type of mode instability, caused by the existence of a mirror-like boundary with non standard boundary conditions.

\section{The Kerr-AdS system}
\label{sec:Kerr-AdS}


We now perform the analogous study for a Kerr BH in asymptotically AdS space. The global structure of AdS is that of a confining box, with a timelike conformal boundary and in this sense conceptually similar to the Kerr-mirror cavity. However, as we shall see, the results appear to differ substantially. In this section, we follow closely the computation in~\cite{Cardoso:2004hs}. Again, to keep the paper self-contained, some details of the computation are provided. 

The Kerr-AdS metric reads:
\begin{align}
\dd s^2 &= - \frac{\Delta_r}{\rho^2} \left(\dd t - \frac{a}{\Sigma} \sin^2 \theta \dd \phi \right)^2
+ \frac{\rho^2}{\Delta_r} \dd r^2 + \frac{\rho^2}{\Delta_{\theta}} \dd \theta^2 \notag \\
&\quad + \frac{\Delta_{\theta}}{\rho^2} \sin^2 \theta \left(a \dd t - \frac{r^2+a^2}{\Sigma} \dd \phi \right)^2 \, ,
\end{align}
with
\begin{gather*}
\Delta_r \equiv (r^2+a^2) \left(1+\frac{r^2}{\ell^2}\right) - 2 M r \, , \qquad \Sigma \equiv 1-\frac{a^2}{\ell^2} \, , \\
\Delta_{\theta} \equiv 1-\frac{a^2}{\ell^2} \cos^2 \theta \, , \qquad \rho^2 \equiv r^2 + a^2 \cos^2 \theta \, ,
\end{gather*}
where $M$ and $J\equiv Ma$ are the BH mass and angular momentum (with $a < \ell$) and $\ell$ is the AdS radius.

Considering the Klein-Gordon equation for a massive scalar field $(\nabla^2 - \mu^2/\ell^2)\Phi = 0$ in this background, with the ansatz
\begin{equation}
\Phi(t,r,\theta,\phi) = e^{-i \omega t + i m \phi} \tilde{S}_l^m(\theta) R(r) \, ,
\end{equation}
where $\tilde{S}_l^m(\theta)$ are the AdS spheroidal angular functions, we obtain the angular equation for $\tilde{S}_l^m(\theta)$,
\begin{multline}
\frac{\Delta_{\theta}}{\sin \theta} \partial_{\theta} \left(\Delta_{\theta} \sin \theta \, \partial_{\theta}\tilde{S}_l^m \right) + \Big[ a^2 \omega^2 \cos^2 \theta - \frac{m^2 \Sigma^2}{\sin^2 \theta} \\
+ A_{lm} \Delta_{\theta} \Big] \tilde{S}_l^m = 0 \, ,
\end{multline}
and the radial equation for $R$,
\begin{multline}
\Delta_r \partial_r \left(\Delta_r \partial_r R \right) + \bigg[ \omega^2 (r^2 + a^2)^2 - 2 M a m \omega r \\
+ a^2 m^2 - \Delta_r \left(a^2 \omega^2 + A_{lm} - \frac{\mu^2}{\ell^2\rho^2} \right) \bigg] R = 0 \, ,
\end{multline}
where $A_{lm}$ is the separation constant.

\subsection{Superradiant modes by a matching method}



We proceed to compute the superradiant modes in a completely parallel way to that used in the Kerr-mirror system.

\subsubsection{Near region $r-r_+ \ll 1/\omega$}


In the near region we can neglect the effects of the cosmological constant and the radial equation reduces to 
%
the same equation as in the near region of the Kerr-mirror \eqref{eq:nearradialeq} and thus the same results apply. In particular, the large $r$ behavior of the ingoing solution in the near region is given by Eq.~\eqref{e:largernear}.
%
%

\subsubsection{Far region $r-r_+ \gg M$}


In the far region the radial equation may be approximated by the wave equation for a massive scalar field mode of frequency $\omega$ and angular quantum number $l$ in AdS spacetime,
\begin{multline}
     (r^2+\ell^2)\frac{\dd^2 R}{\dd r^2} +2 \left(2r + \frac{\ell^2}{r}\right) \frac{\dd R}{\dd r} \\
     + \ell^2\left[\frac{\omega^2\ell^2}{\ell^2+r^2} - \frac{l(l+1)}{r^2} - \mu^2 \right] R = 0 \, .
\end{multline}

Introducing a new radial coordinate
\begin{equation}
	x \equiv 1 + \frac{r^2}{\ell^2} \, , \quad 1 \leqslant x < \infty \, ,
\end{equation}
and defining
\begin{equation}
	R(x) \equiv x^{\frac{\omega\ell}{2}} (1-x)^{\frac{l}{2}} F(x) \, ,
\end{equation}
the radial equation can be written as
\begin{equation}
	x(1-x) \frac{\dd^2 F}{\dd x^2} + \left[\gamma - (\alpha+\beta+1)x\right] \frac{\dd F}{\dd x} - \alpha \beta F = 0 \, ,
\end{equation}
with
\begin{subequations}
\begin{align}
	\alpha &\equiv \frac{\omega\ell}{2} + \frac{l}{2} + \frac{3}{4} + \frac{1}{4}\sqrt{9+4\mu^2} \, , \\
	\beta & \equiv \frac{\omega\ell}{2} + \frac{l}{2} + \frac{3}{4} - \frac{1}{4}\sqrt{9+4\mu^2} \, , \\
	\gamma & \equiv 1 + \omega\ell \, .
\end{align}
\end{subequations}
This is the Gaussian hypergeometric equation \cite{NIST} and the most general solution for $R$ may be written as
\begin{align}
	R(x) &= (1-x)^{\frac{l}{2}} \left[ C x^{\frac{\omega\ell}{2}-\alpha}  F\left(\alpha, \alpha-\gamma+1; \alpha-\beta+1; \tfrac{1}{x}\right) \right. \notag \\
	&\quad \left. + D x^{\frac{\omega\ell}{2}-\beta} F\left(\beta, \beta-\gamma+1; \beta-\alpha+1; \tfrac{1}{x}\right) \right] \, .
\end{align}
$C,D$ are two integration constants. As $x \to \infty$, 
\begin{equation}
R(x) \sim (-1)^{\frac{l}{2}} \left(C x^{-\frac{3}{4} - \frac{1}{4}\sqrt{9+4\mu^2}} + D x^{-\frac{3}{4} + \frac{1}{4}\sqrt{9+4\mu^2}} \right) \ ,
\end{equation}
and hence imposing DBCs is equivalent to setting $D=0$. Herein, we wish to consider the more generic case of  RBCs, that is,
\begin{equation} \label{eq:RobinBCAdS}
	C \sin(\zeta) = D \cos(\zeta) \, , \quad \zeta \in [0,\pi) \, ,
\end{equation} 
which may be imposed when $-\frac{9}{4} < \mu^2 < - \frac{5}{4}$.

The small $\omega r$ behavior of this solution may be obtained by making the transformation $\frac{1}{x} \to 1-x$ \cite{NIST},
\begin{align*}
    R(x) &= x^{\frac{\omega\ell}{2}} (1-x)^{\frac{l}{2}} \, \Big[ E_1 \, x^{1-\gamma} (x-1)^{\gamma-\alpha-\beta} \\
    &\qquad \times F(1-\alpha,1-\beta;\gamma-\alpha-\beta+1;1-x) \\
    &\quad + E_2 \, F(\alpha,\beta;\alpha+\beta-\gamma+1;1-x) \Big] \, ,
\end{align*}
where
\begin{align*}
    E_1 &\equiv \Gamma(\alpha+\beta-\gamma) \left(C \tfrac{\Gamma(\alpha-\beta+1)}{\Gamma(\alpha)\Gamma(\alpha-\gamma+1)} + D \tfrac{\Gamma(\beta-\alpha+1)}{\Gamma(\beta)\Gamma(\beta-\gamma+1)} \right) \, , \\
    E_2 &\equiv \Gamma(\gamma-\alpha-\beta) \left(C \tfrac{\Gamma(\alpha-\beta+1)}{\Gamma(1-\beta)\Gamma(\gamma-\beta)} + D \tfrac{\Gamma(\beta-\alpha+1)}{\Gamma(1-\alpha)\Gamma(\gamma-\alpha)} \right) \, 
\end{align*}
and \eqref{eq:RobinBCAdS} is satisfied.
As $x \to 1$ and $x-1 \to r^2/\ell^2$, one has
\begin{equation} \label{eq:smallrfarAdS}
	R(r) \approx (-1)^{\frac{l}{2}} \left[E_1 \, \ell^{l+1} \, r^{-l-1} + E_2 \, \ell^{-l} \, r^l \right] \, .
\end{equation}
%

\subsubsection{Matching in overlapping region $M \ll r-r_+ \ll 1/\omega$}

In the overlapping region $M \ll r-r_+ \ll 1/\omega$ one can match the large $r$ solution in the near region \eqref{e:largernear} with the small $r$ solution in the far region \eqref{eq:smallrfarAdS}, giving
\begin{align} \label{eq:matchAdS}
	\frac{E_2}{E_1} &= \frac{(r_+-r_-)^{2l+1} \Gamma(l+1)\Gamma(-2l-1)}{\Gamma(2l+1)\Gamma(-l)} \notag \\
	&\quad \times \frac{\Gamma(l+1-i2\varpi)}{\Gamma(-l-i2\varpi)} \, ,
\end{align}
where \eqref{eq:RobinBCAdS} is satisfied.

\subsection{Computation of the quasinormal frequencies}

%
\begin{figure*}[t]
	\centering
	\includegraphics[width=0.47\linewidth]{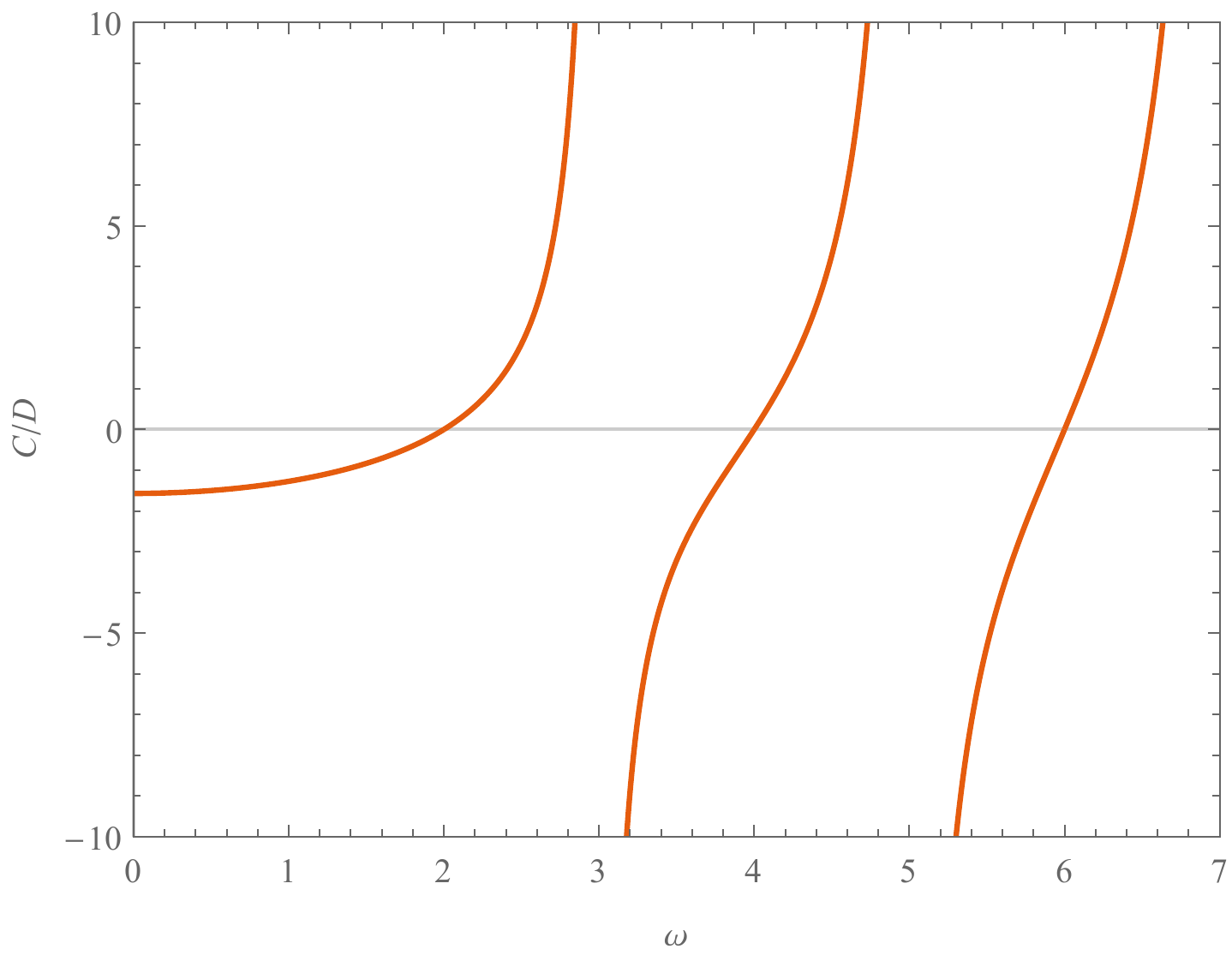} \hspace*{5ex}
	\includegraphics[width=0.47\linewidth]{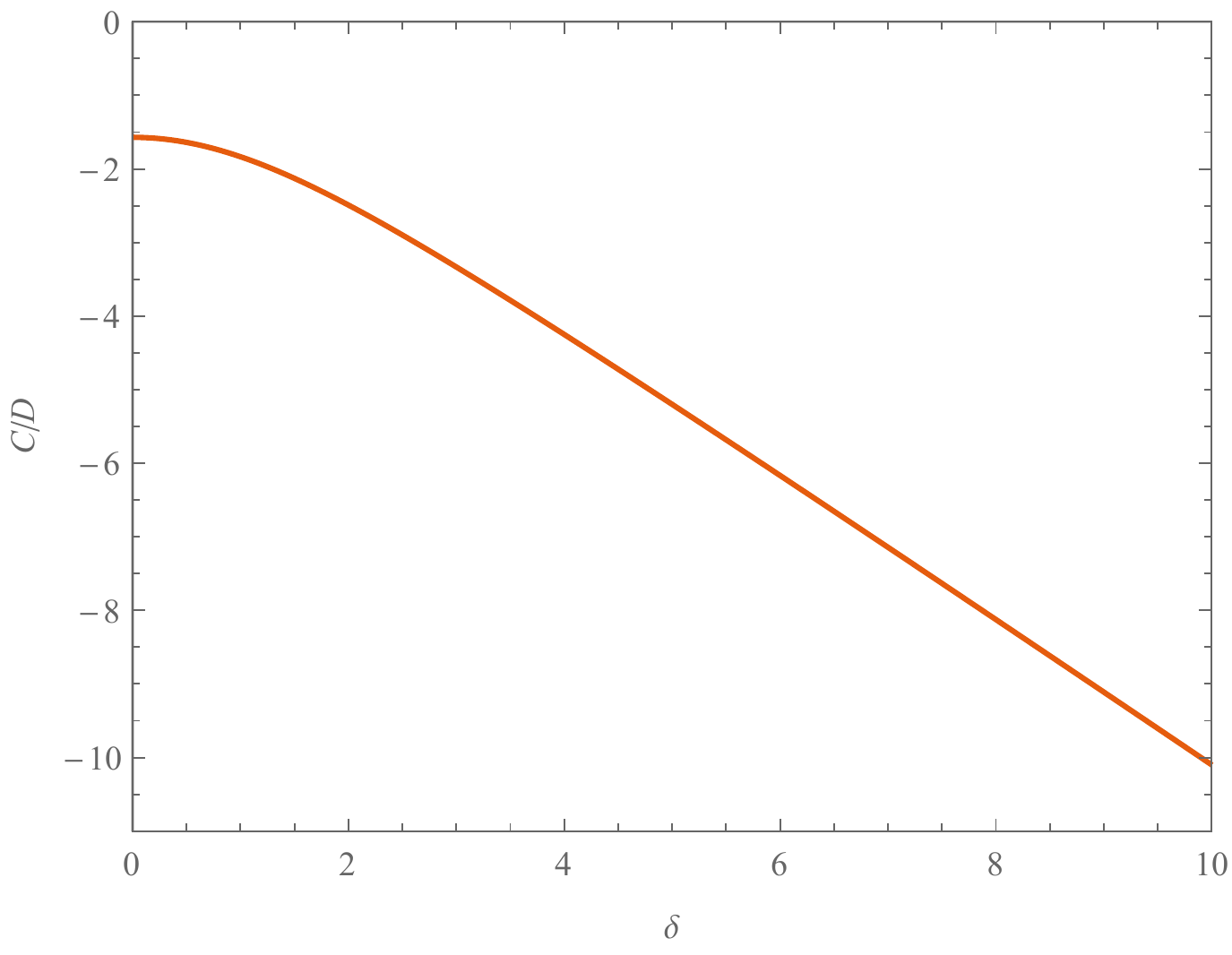}
	\caption{\label{fig:plot-Kerr-AdS-CD}Normal frequencies in pure AdS (with $\ell=1$). The plots show $C/D$, given by \eqref{eq:CD-KerrAdS}, $vs.$ $\omega$ (left panel) and $vs.$ $\delta$ (right panel) for a mode with $\mu^2=-2$ and $l=1$. The real zeros of this function correspond to the Neumann normal frequencies, whereas the singularities correspond to the Dirichlet normal frequencies. The purely imaginary frequencies never occur for either Dirichlet or Neumann boundary conditions.
	\vspace*{1ex}
	}
\end{figure*}

The strategy to compute the quasinormal frequencies in the Kerr-AdS BH is, again, to treat them as a small ``perturbation'' to the \emph{normal} frequencies of pure AdS, which is justified for Kerr-AdS BHs with $r_+ \ll \ell$ and $a \ll \ell$. 

The normal frequencies are determined by requiring RBCs at infinity, as in \eqref{eq:RobinBCAdS}, and regularity at the origin, which implies that $E_1 = 0$, $i.e$
\begin{align} \label{eq:CD-KerrAdS}
\frac{D}{C} = - \frac{\Gamma(\alpha-\beta+1)\Gamma(\beta)\Gamma(\beta-\gamma+1)}{\Gamma(\beta-\alpha+1)\Gamma(\alpha)\Gamma(\alpha-\gamma+1)}  \, .
\end{align}
Plots of $C/D$ as a function of real frequency $\omega$ and purely imaginary frequency $i\delta$ are shown in Fig.~\ref{fig:plot-Kerr-AdS-CD}.

Hence, in order for the RBCs \eqref{eq:RobinBCAdS} to be satisfied one has that, for $\zeta \neq \frac{\pi}{2}$,
\begin{equation}
\tan(\zeta) = - \frac{\Gamma(\alpha-\beta+1)\Gamma(\beta)\Gamma(\beta-\gamma+1)}{\Gamma(\beta-\alpha+1)\Gamma(\alpha)\Gamma(\alpha-\gamma+1)}  \, .
\end{equation}
For fixed $l$ and $\zeta$, this gives a discrete spectrum of normal frequencies. In the case of DBCs, $\zeta=0$, they are given by
\begin{equation}
\omega_n^{\rm (D)} = \pm \frac{1}{\ell}\left(l+\frac{3}{2}+\frac{1}{2}\sqrt{9+4\mu^2}+2n \right) \, , \quad n \in \bN \, ,
\end{equation}
and for Neumann boundary conditions, $\zeta=\frac{\pi}{2}$, they are
\begin{equation}
\omega_n^{\rm (N)} = \pm \frac{1}{\ell}\left(l+\frac{3}{2}-\frac{1}{2}\sqrt{9+4\mu^2}+2n \right) \, , \quad n \in \bN \, .
\end{equation}
For other RBCs, the frequencies, including the purely imaginary ones, need to be computed numerically. 
Without loss of generality, we only focus on positive real part and imaginary part of the frequencies.

We note that the real normal frequency tends to zero as it approaches a critical value denoted $\zeta_{\rm AdS}$, given by
\begin{equation}
\zeta_{\rm AdS} \equiv \arctan(-\frac{\Gamma\left(1+\tilde{\mu}\right) \Gamma\left(\frac{l}{2}+\frac{3}{4}-\frac{1}{2}\tilde{\mu}\right)^2}{\Gamma\left(1-\tilde{\mu}\right) \Gamma\left(\frac{l}{2}+\frac{3}{4}+\frac{1}{2}\tilde{\mu}\right)^2}) \, ,
\end{equation}
with $\tilde{\mu} \equiv \frac{1}{2}\sqrt{9+4\mu^2}$. Then, for $\zeta \in (\zeta_{\rm AdS},\pi)$, there is a normal frequency which is purely imaginary and which diverges as $\zeta \to \pi^-$, that is, when $C/D \to - \infty$. This is an intrinsic, unstable mode due to the AdS asymptotics, which was also analyzed in~\cite{Dappiaggi:2016fwc} in pure AdS in all spacetime dimensions.

%
%
%

\subsubsection{Approximation $\Im[\omega] \ll \Re[\omega]$}

In the case of a small, slowly rotating Kerr-AdS BH, the quasinormal frequencies will be approximately equal to those of pure AdS, with a small imaginary part in the case of the real normal frequencies, or a small real part in the case of the purely imaginary normal frequencies. In this subsection, we take the real part of the quasinormal frequencies to be equal to the corresponding real AdS normal frequencies, and compute their imaginary part assuming that $\Im[\omega] \ll \Re[\omega]$, that is, write the full quasinormal frequencies as $\omega_{\rm QN} = \omega_n + i \delta$, with $\delta \ll \omega_n$. By replacing it in \eqref{eq:matchAdS} and solving for $\delta$, one obtains
\begin{align}
\delta = 
i \frac{2^{2l+2} (l!)^2 \, (r_+-r_-)^{2 l+1}}{\ell^{2(l+1)}\left[(2l+1)!! (2l)!\right]^2}
\frac{\Gamma (l-2 i \varpi +1)}{\Gamma (-l-2 i \varpi )} \frac{\Sigma_1}{\Sigma_2}
 \, ,
\end{align}
where 
\begin{align*}
\Sigma_1 &\equiv \frac{\Gamma(\alpha) \Gamma(\alpha-\gamma+1)}{\Gamma(1-\beta)\Gamma(\gamma-\beta)} 
- \frac{\Gamma(\beta) \Gamma(\beta-\gamma+1)}{\Gamma(1-\alpha)\Gamma(\gamma-\alpha)}  \, , \\
\Sigma_2 &\equiv \psi(\alpha) - \psi(\alpha-\gamma+1) - \psi(\beta) + \psi(\beta-\gamma+1) \, ,
\end{align*}
and $\psi(z) = \Gamma'(z)/\Gamma(z)$ is the digamma function. 

\begin{figure*}[t!]
	\centering
	\includegraphics[width=0.36\linewidth]{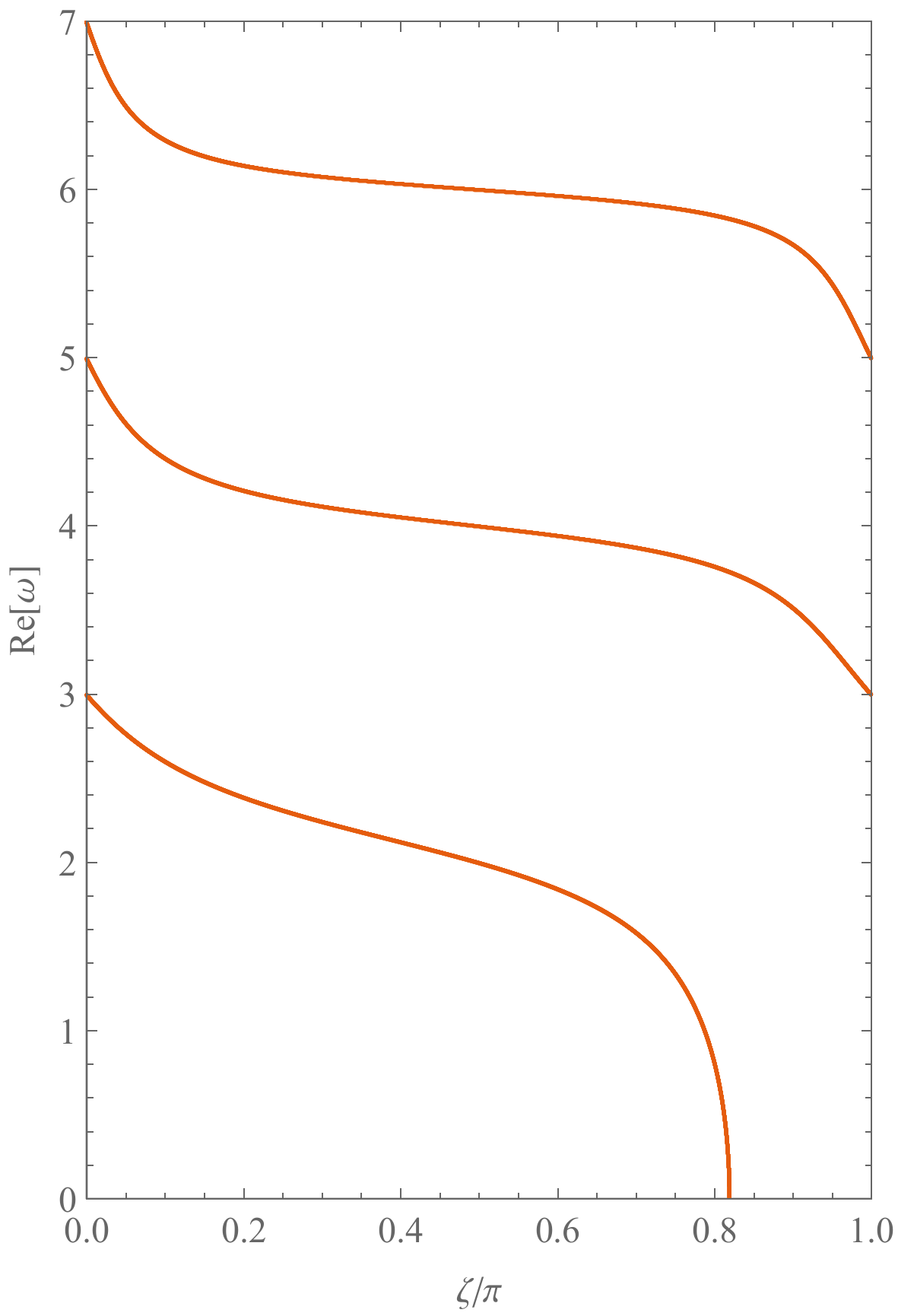} \hspace{10ex}
	\includegraphics[width=0.37\linewidth]{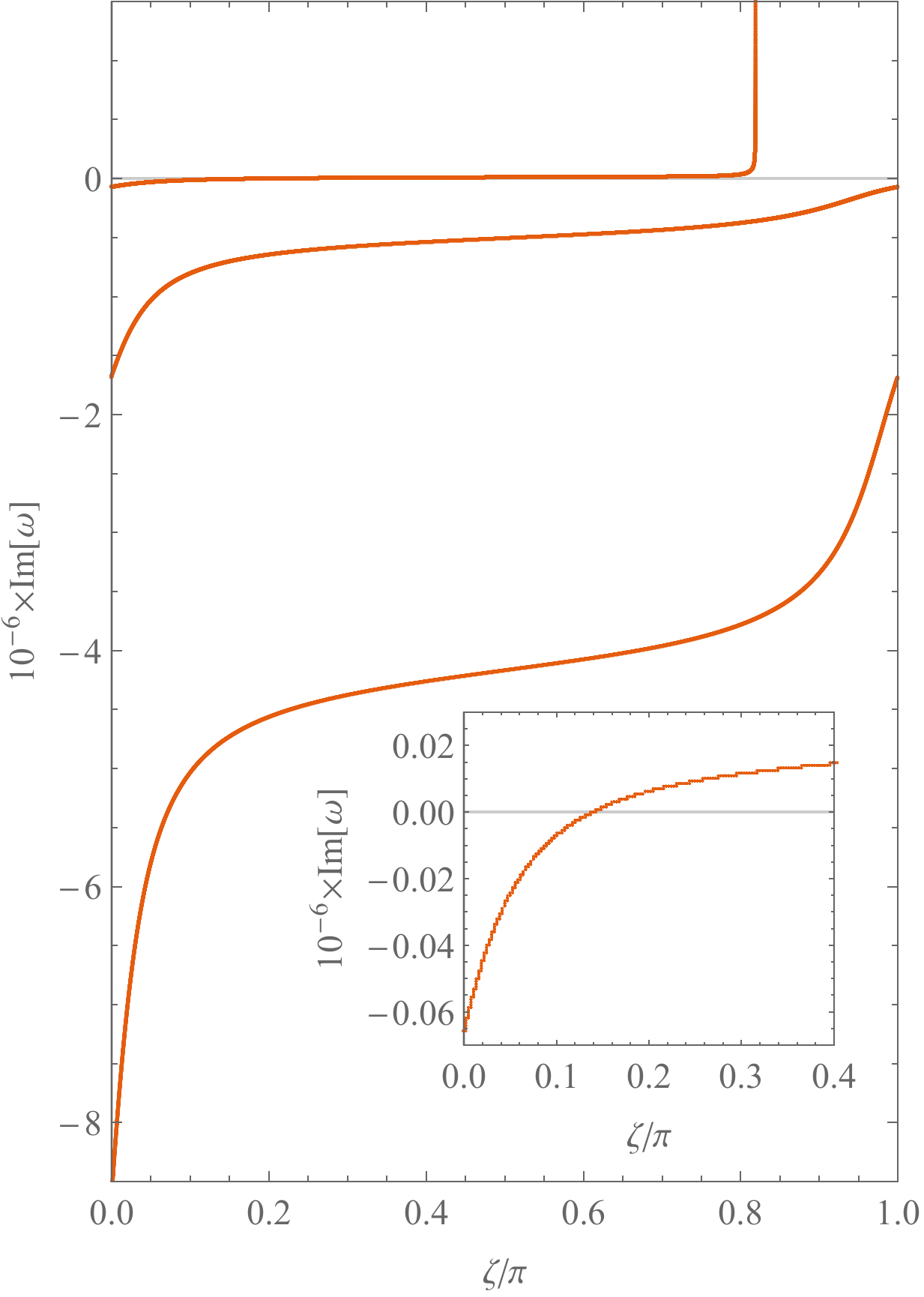}
	\caption{\label{fig:plot-Kerr-AdS-1}Quasinormal frequencies in Kerr-AdS close to the real normal frequencies. We show the real part (which coincides with the normal frequencies) and the imaginary part of the frequencies $vs.$ $\zeta/\pi$ for the Kerr-AdS system with $M=0.01$, $a=0.001$, and a mode with $\mu^2=-2$, $l=1$ and $m=1$ ($m \Omega_{\mathcal{H}} \approx 2.506$), computed in the regime $\Im[\omega] \ll \Re[\omega]$. In this case, the Dirichlet frequencies have real part 4, 6, ..., while the Neumann have real part 1, 3, 5, ...
	\vspace*{4ex}
	}
\end{figure*}

For DBCs, $\omega_n = \omega_n^{\rm (D)}$,
\begin{align*}
\delta^{\rm (D)} &= i (-1)^{3l+1} \frac{\pi \ell^{-2(l+1)} (r_+-r_-)^{2l+1} \Gamma\left(n+l+\frac{3}{2}\right)}{2^{4l} n! \Gamma\left(l+\frac{1}{2}\right)^2 \Gamma\left(l+\frac{3}{2}\right)^2} \notag \\
&\quad \times \frac{\Gamma \left(n+l+\frac{3}{2}+\frac{1}{2} \sqrt{4 \mu ^2+9}\right)}{\Gamma \left(n +1+\frac{1}{2} \sqrt{4 \mu ^2+9} \right)} \frac{\Gamma (l-2 i \varpi +1)}{\Gamma (-l-2 i \varpi)} \, ,
\end{align*}
and for NBCs, $\omega_n = \omega_n^{\rm (N)}$,
\begin{align*}
\delta^{\rm (N)} &= i (-1)^{3l+1} \frac{\pi \ell^{-2(l+1)}(r_+-r_-)^{2l+1} (n+2l+1)!!}{2^{n+4l} n! (2l+1)!! \Gamma\left(l+\frac{1}{2}\right)^2 \Gamma\left(l+\frac{3}{2}\right)} \notag \\
&\quad \times \frac{\Gamma \left(n+l+\frac{3}{2}-\frac{1}{2} \sqrt{4 \mu ^2+9}\right)}{\Gamma \left(n +1-\frac{1}{2} \sqrt{4 \mu ^2+9} \right)} \frac{\Gamma (l-2 i \varpi +1)}{\Gamma (-l-2 i \varpi)} \, .
\end{align*}
We can simplify the above expressions using the relation $\Gamma(z+1)=z\Gamma(z)$,
%
%
obtaining
\begin{align}
\delta^{\rm (D)} &= - \frac{\pi \ell^{-2(l+1)} (r_+-r_-)^{2l+1} \Gamma\left(n+l+\frac{3}{2}\right)}{2^{4l} n! \Gamma\left(l+\frac{1}{2}\right)^2 \Gamma\left(l+\frac{3}{2}\right)^2} \notag \\
&\quad \times \frac{\Gamma \left(n+l+\frac{3}{2}+\frac{1}{2} \sqrt{4 \mu ^2+9}\right)}{\Gamma \left(n +1+\frac{1}{2} \sqrt{4 \mu ^2+9} \right)} \, 2\varpi \prod_{k=1}^{l}\left(k^2+4\varpi^2\right) \, ,
\end{align}
and
\begin{align}
\delta^{\rm (N)} &= - \frac{\pi \ell^{-2(l+1)} (r_+-r_-)^{2l+1} (n+2l+1)!!}{2^{n+4l} n! (2l+1)!! \Gamma\left(l+\frac{1}{2}\right)^2 \Gamma\left(l+\frac{3}{2}\right)} \notag \\
&\quad \times \frac{\Gamma \left(n+l+\frac{3}{2}-\frac{1}{2} \sqrt{4 \mu ^2+9}\right)}{\Gamma \left(n +1-\frac{1}{2} \sqrt{4 \mu ^2+9} \right)} \, 2\varpi \prod_{k=1}^{l}\left(k^2+4\varpi^2\right) \, .
\end{align}

For DBCs (the case studied in \cite{Cardoso:2004hs}) and NBCs, when $\varpi < 0$, $\delta^{\rm (D)} > 0$ and  $\delta^{\rm (N)} > 0$, for all $n \in \mathbb{N}$. These are the expected results when the superradiance condition $0 < \Re[\omega_n] < m \Omega_{\mathcal{H}}$ holds. 

For other RBCs, the real and imaginary part of the frequencies need to be obtained numerically. In Fig.~\ref{fig:plot-Kerr-AdS-1} we show the real and imaginary part of the frequencies for a representative example. 

Note that the numerically obtained $\delta$ appears to diverge when $\zeta \to \zeta_{\rm AdS}$. This happens when the corresponding real part of the frequency, which we took to be equal to the AdS normal frequency, tends to zero, and the assumption $\delta \ll \omega$ is not valid anymore. Therefore, the numerical results we obtained are only valid when the real part is sufficiently large.

\subsubsection{Approximation $\Re[\omega] \ll \Im[\omega]$}

\begin{figure*}[t!]
	\centering
	\includegraphics[width=0.36\linewidth]{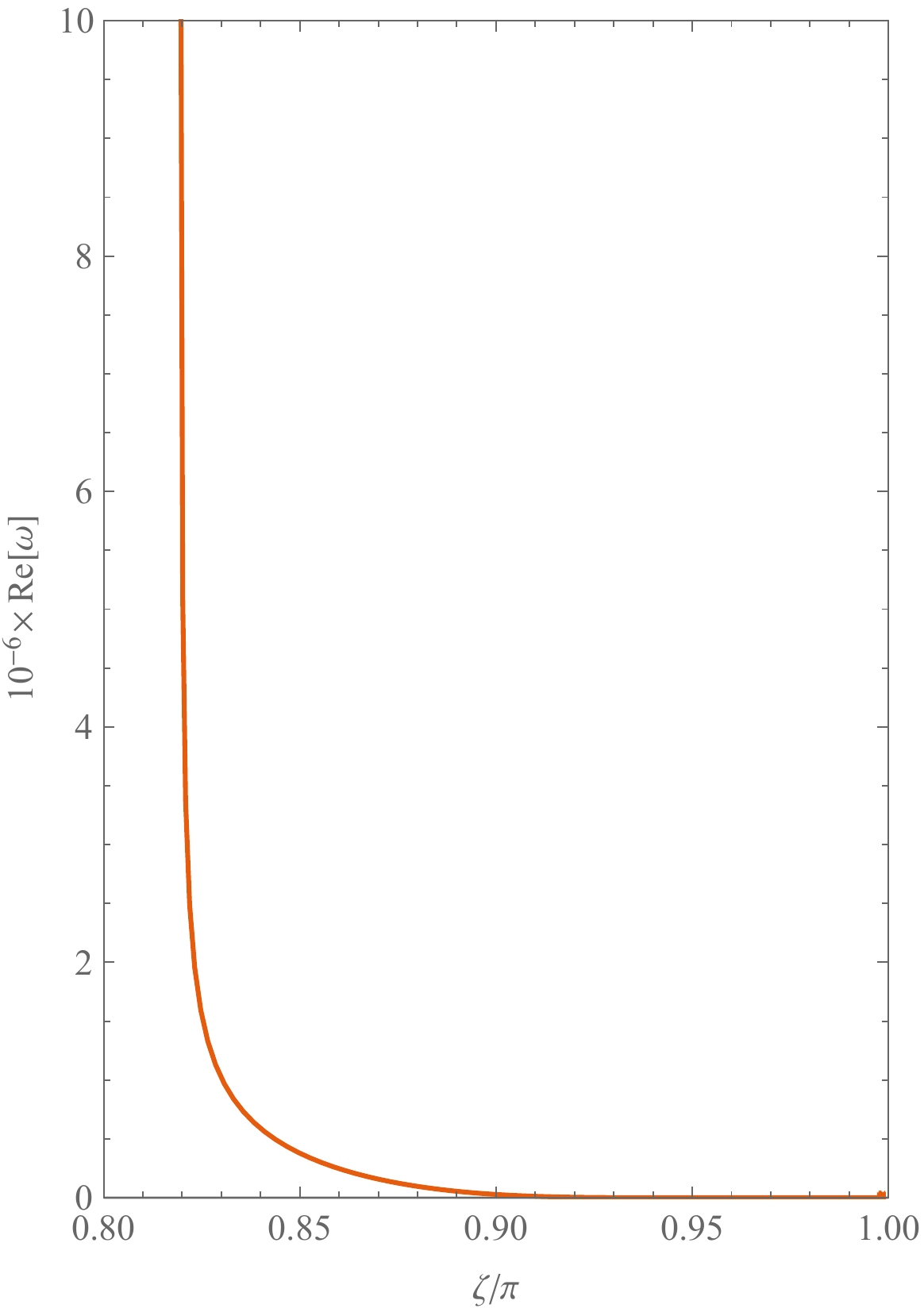} \hspace{10ex}
	\includegraphics[width=0.37\linewidth]{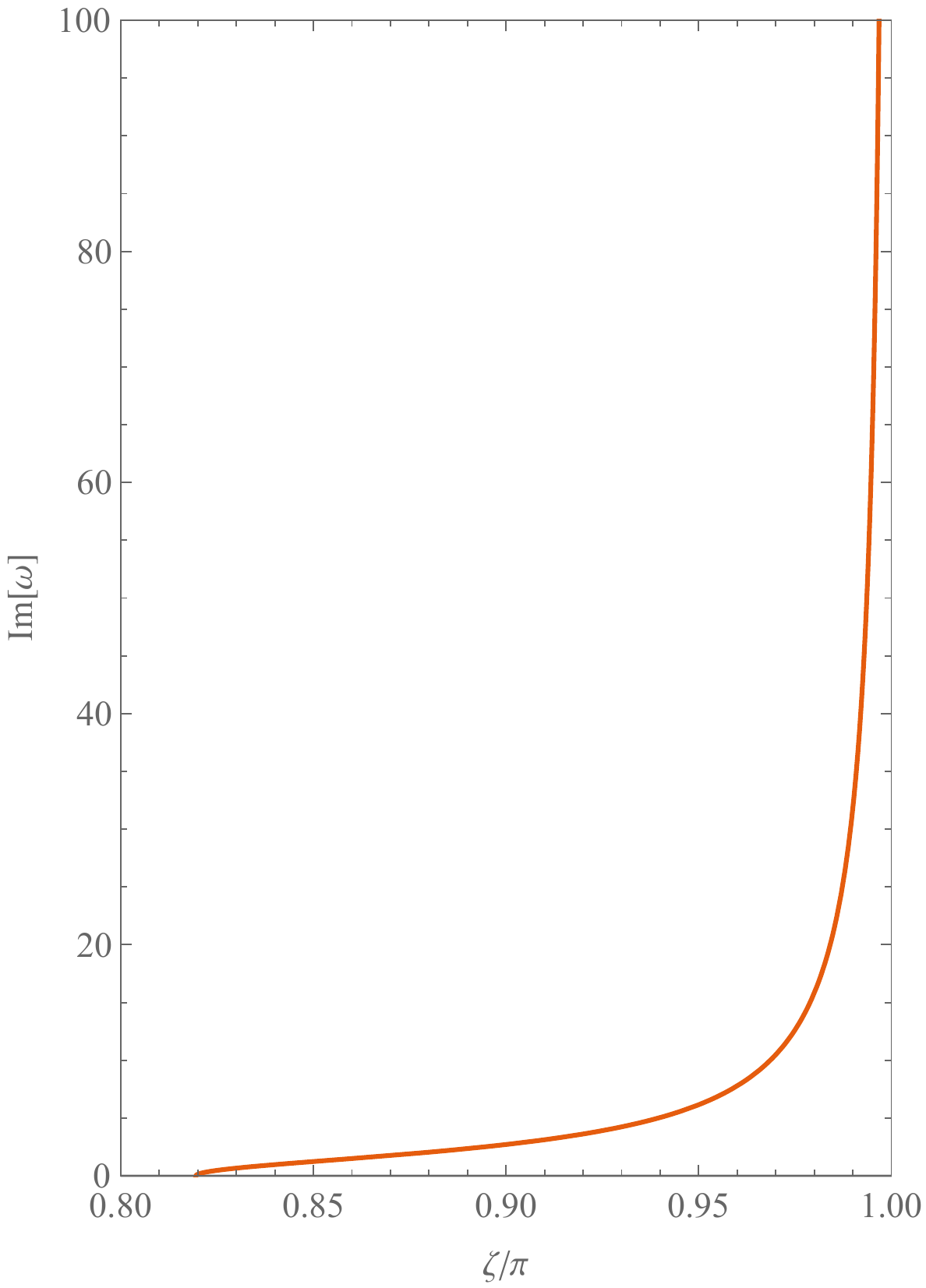}
	\caption{\label{fig:plot-Kerr-AdS-2}Quasinormal frequencies in Kerr-AdS close to the imaginary ``normal" frequencies. We show the real part and the imaginary part (which coincides with the ``normal" frequencies) of the frequencies $vs.$ $\zeta/\pi$ for the Kerr-AdS system with $M=0.01$, $a=0.001$, and a mode with $\mu^2=-2$, $l=1$ and $m=1$ ($m \Omega_{\mathcal{H}} \approx 2.506$), computed in the regime $\Re[\omega] \ll \Im[\omega]$. The numerical results are not valid for $\zeta \lesssim 0.85\pi$.
	\vspace*{2ex}
	}
\end{figure*}

In this subsection, we compute the quasinormal frequencies whose imaginary part is much greater than the real part, by taking the imaginary part to be the same as the corresponding purely imaginary AdS normal frequencies and by computing the respective real part, such that $\omega_{\rm QN} = \omega + i \delta$, with $\omega \ll \delta$. This can be obtained by using \eqref{eq:matchAdS}. Since the purely imaginary AdS normal frequencies only exist for $\zeta \in (\zeta_{\rm AdS}, \pi)$, the corresponding BH real part have to be obtained numerically and in Fig.~\ref{fig:plot-Kerr-AdS-2} we show a representative example.

Observe that, as $\zeta \to \pi^-$, the real part $\omega \to 0$ and the imaginary part $\delta \to \infty$.
This behavior is identical to the one found in the case of the (2+1)-dimensional BTZ BH in \cite{Dappiaggi:2017pbe}. As explained therein, the flux of energy across the horizon for modes with such frequencies is directed \emph{towards} the BH, since the flux of energy may be shown to be proportional to\footnote{Note that, for $\delta \ll \omega$, the sign of the flux of energy across the horizon is essentially determined by $\omega-m\Omega_{\mathcal{H}}$, as expected.} $\delta^2 + \omega(\omega-m\Omega_{\mathcal{H}})$. However, angular momentum is still being extracted from the BH (as its flux across the horizon is proportional to $\omega-m\Omega_{\mathcal{H}}$), which shows that superradiance is still occurring. This is an intrinsic, mode instability which occurs when a subset of Robin boundary conditions are imposed at the AdS boundary, as such a growing mode is also present in pure AdS${}_4$ for a massive scalar field with certain RBCs at infinity \cite{Dappiaggi:2016fwc}. In this regime, this mode instability is dominant in comparison with the superradiant instability and, therefore, energy flows into the BH. 

Note that the numerical results show that as $\zeta \to \zeta_{\rm AdS}$ the real part $\omega$ appears to diverge while the imaginary part $\delta$ tends to zero, meaning that the assumption $\omega \ll \delta$ is not valid anymore and the results stop being reliable.


\section{Conclusions}
\label{sec:conclusions}

In this paper we discuss the nature of superradiant instabilities of the Kerr-mirror and Kerr-AdS BH triggered by scalar fields when RBCs are imposed at either the mirror's location or the AdS infinity. Our results show that the two systems present the usual superradiant instability and a new type of mode instability for certain RBCs.

\begin{itemize}
\item In the case of the Kerr-mirror system, the superradiant modes have similar qualitative behaviors for general RBCs as for the DBCs studied in \cite{Cardoso:2004nk}. Namely, the superradiant modes have $\Im[\omega] > 0$ when $0<\Re[\omega]<m\Omega_{\mathcal{H}}$. However, our calculations show further that, for a certain range of RBCs, there exists a mode with $\Im[\omega] > 0$ much greater than $\Re[\omega]$ and whose flux of energy across the horizon is directed \emph{towards} the BH. This is a new type of mode instability which is caused when certain RBCs are imposed at the mirror.
\item In the case of Kerr-AdS BH, we also have modes with $\Im[\omega] > 0$ when $0<\Re[\omega]<m\Omega_{\mathcal{H}}$. However, as we consider RBCs for which the $\Re[\omega]$ approaches zero, we have numerical evidence that $\Im[\omega]$ becomes arbitrarily large and, as a consequence, the flux of energy of these modes across the horizon is directed \emph{towards} the BH. Angular momentum is still being extracted from the BH, showing that superradiance is still occurring, but it is subdominant in comparison with the mode instability caused when these RBCs are imposed on the AdS boundary.
\end{itemize}

It remains to study these superradiant modes in the Kerr-mirror and the Kerr-AdS systems without the technical restrictions in place in this paper, namely considering BHs of arbitrary size and rotation speed, and using numerical methods to compute frequencies with arbitrary real to imaginary part ratio. This would allow us to continuously connect between the two asymptotic regimes explored in this paper.

Another interesting question arises from the existence of the new unstable mode when certain RBCs are imposed at the boundaries of these confined spacetimes. In the simplest of these cases, one can consider pure Minkowski spacetime with a mirror-like boundary at a finite radial coordinate at which those RBCs are imposed. Given the new unstable mode, one may wonder what is the final state of the system, whose determination requires to go beyond the linear theory explored in this paper. One possibility is that, after taking into account the backreaction, the system stabilizes into a state with negative energy. The second, more drastic possibility is that the system collapses has no ground state and, therefore, the field theory is pathological. We hope to return to this question in the future.


\begin{acknowledgments}
We would like to thank R.~Brito, C.~Dappiaggi and B.~A.~Juarez-Aubry for useful discussions.
The work of H.~F.\ was supported by the INFN postdoctoral fellowship ``Geometrical Methods in Quantum Field Theories and Applications'', and in part by a fellowship of the ``Progetto Giovani GNFM 2017 -- Wave propagation on lorentzian manifolds with boundaries and applications to algebraic QFT'' fostered by the National Group of Mathematical Physics (GNFM-INdAM). 
C.~H. acknowledge funding from the FCT-IF programme.  This  work  was  partially supported by the H2020-MSCA-RISE-2015 Grant No.   StronGrHEP-690904, the H2020-MSCA-RISE-2017 Grant No.   FunFiCO-777740  and  by  the  CIDMA  project UID/MAT/04106/2013. We  would also  like  to  acknowledge networking support by the COST Action GWverse
CA16104.

\end{acknowledgments}


\end{document}